\documentclass[a4paper,fleqn,usenatbib]{mnras}

\usepackage{newtxtext,newtxmath}

\usepackage[T1]{fontenc}
\usepackage{ae,aecompl}

\usepackage{graphicx}	
\usepackage{amsmath}	
\usepackage{amssymb}	
\usepackage{pdflscape}

\title[Iron lines in SDO/EVE data as density diagnostics]{An assessment of Fe\,{\Large \textbf{XX}} -- 
Fe\,{\Large \textbf{XXII}} emission lines in SDO/EVE data as diagnostics for high density solar flare plasmas using EUVE stellar observations}

\author[F. P. Keenan et al.]{
F. P. Keenan,$^{1}$\thanks{E-mail: f.keenan@qub.ac.uk}
R. O. Milligan,$^{1,2,3}$
M. Mathioudakis$^{1}$
and D. J. Christian$^{4}$
\\
$^{1}$Astrophysics Research Centre, School of Mathematics and Physics, Queen's University Belfast, BT7 1NN, UK\\
$^{2}$Department of Physics, Catholic University of America, 620 Michigan Ave. NE, Washington DC 20064, USA\\
$^{3}$Solar Physics Laboratory (Code 671), Heliophysics Science Division, NASA Goddard Space Flight Center, Greenbelt, MD 20771, USA\\
$^{4}$Department of Physics and Astronomy, California State University, Northridge, CA 91330, USA
}

\date{Accepted 2017 February 27. Received 2017 February 7; in original form 2016 December 22.}

\pubyear{2016}

\begin{document}
\label{firstpage}
\pagerange{\pageref{firstpage}--\pageref{lastpage}}
\maketitle

\begin{abstract}
The Extreme Ultraviolet Variability Experiment (EVE) on the Solar Dynamics Observatory obtains extreme-ultraviolet (EUV) spectra of the full-disk Sun at a spectral resolution of $\sim$\,1\,\AA\ and cadence of 10\,s. Such a spectral resolution would normally be considered to be too low for the reliable determination of electron density (N$_{e}$) sensitive emission line intensity ratios, due to blending.
However, previous work has shown that a limited number of Fe\,{\sc xxi} features in the 90--160\,\AA\ wavelength region of EVE do provide useful 
N$_{e}$-diagnostics at relatively low flare densities (N$_{e}$ $\simeq$ 
10$^{11}$--10$^{12}$\,cm$^{-3}$).  Here we investigate if additional highly ionised Fe line ratios in the EVE 90--160\,\AA\ range 
may be reliably employed as N$_{e}$-diagnostics. In particular, the potential for such diagnostics to 
provide density estimates for high N$_{e}$ ($\sim$\,10$^{13}$\,cm$^{-3}$) flare plasmas is assessed. 
Our study employs EVE spectra for X-class flares, combined with observations of highly active late-type stars from the Extreme Ultraviolet Explorer (EUVE)
satellite 
plus experimental data for well-diagnosed tokamak plasmas, both of which are similar in wavelength coverage and spectral resolution to those from EVE. Several ratios are identified in EVE data
which yield consistent values of electron density, including Fe\,{\sc xx}
113.35/121.85 and Fe\,{\sc xxii} 114.41/135.79, with confidence in their reliability as N$_{e}$-diagnostics provided by the EUVE and tokamak results. These ratios also allow
the determination of density in solar flare plasmas up to values of $\sim$\,10$^{13}$\,cm$^{-3}$.
\end{abstract}




\section{Introduction}

Electron density (N$_{e}$) is a fundamental physical parameter of a plasma, and one of the main ways for determining this quantity in remote astrophysical sources is via emission line intensity ratios which are sensitive to variations in N$_{e}$. In the 
case of the high temperature solar transition region and corona, many of the emission lines lie at extreme-ultraviolet (EUV)
wavelengths ($\sim$\,100--1200\,\AA), and for more than half a century numerous researchers have worked on the calculation
of theoretical line ratio N$_{e}$-diagnostics involving EUV transitions (see, for example, \citealt{jordan66}; \citealt{keenan96};
\citealt{delzannabadnell16}).
Such diagnostics have generally been used in conjunction with medium to high spectral resolution 
solar observations obtained with rocket-borne or satellite-based instrumentation, required to adequately resolve and hence 
reliably measure the relevant emission lines. Examples include the S082A instrument on board the Skylab space station, which 
had a resolution of $\sim$\,0.1\,\AA\ \citep{dere78}, the Solar EUV Rocket Telescope and Spectrograph (SERTS) at 0.05--0.08\,\AA\
resolution \citep{thomasneupert94}, the Coronal Diagnostic Spectrometer (CDS) on the Solar and Heliospheric Observatory (SOHO) satellite at $\sim$\,0.4\,\AA\
resolution \citep{harrison97} and the EUV Imaging Spectrometer (EIS) on the Hinode satellite at $\sim$\,0.07\,\AA\
resolution \citep{young07}.

The Solar Dynamics Observatory (SDO), launched on 2010 February 11, contains a number of instruments including the Extreme
Ultraviolet Variability Experiment (EVE), which obtains a full-disk EUV spectrum every 10\,s using its MEGS-A component, albeit
at a relatively low resolution of $\sim$\,1\,\AA\ \citep{woods11}. Although the EVE temporal resolution is unprecedented for
an EUV spectrometer, allowing detailed studies of time-dependent phenomena such as flares, one would expect the spectral resolution to
be too low to provide reliable electron density diagnostics. However, \citet{milligan12} found that a few Fe\,{\sc xxi}
emission line ratios in the 90--160\,\AA\ portion of the MEGS-A flare spectra provide useful electron density diagnostics for values of 
N$_{e}$ in the range $\sim$\,10$^{11}$--10$^{12}$\,cm$^{-3}$, hence allowing the study of the temporal 
evolution of relatively low flare density.

The short-wavelength spectrometer on the Extreme Ultraviolet Explorer (EUVE) satellite obtained spectra of astrophysical sources in the
70--190\,\AA\ region at a resolution of $\sim$\,0.5\,\AA\ \citep{abbott96}. It observed many highly-active late-type stars (see, for
example, \citealt{craig97}), and due to the comparable spectral coverage 
and resolution with EVE flare data, the EUVE and EVE 
observations can appear remarkably similar in terms of the presence of highly ionised Fe lines, as illustrated in Figure 1. 
Hence the EUVE spectra may be employed as a `testbed' for Fe ion line ratio diagnostics in EVE observations, and in particular allow us to assess the potential usefulness of such diagnostics for high N$_{e}$ flare plasmas, as active late-type stars can contain 
high density (up to $\sim$\,10$^{13}$\,cm$^{-3}$ or greater) coronal material.

\begin{figure*}
		\includegraphics[width=16cm]{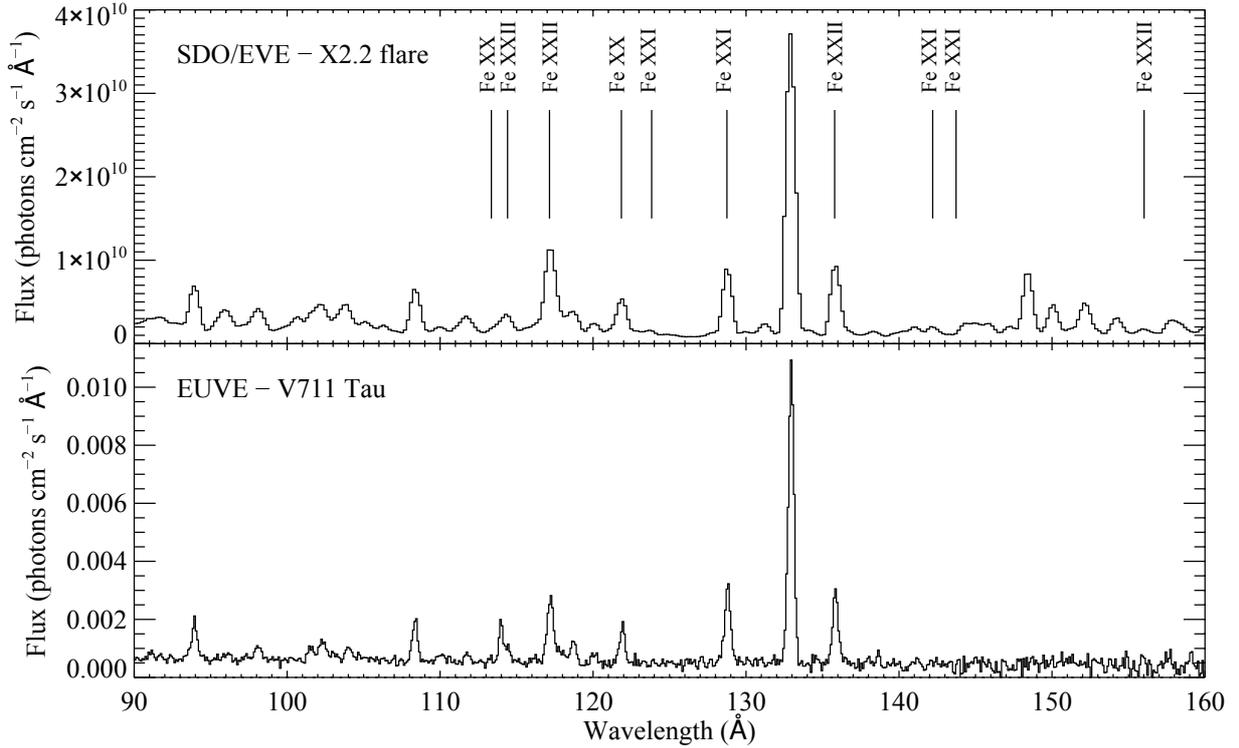}
    \caption{Comparison of the SDO/EVE spectrum of the solar flare of 2011 February 15 at 01:55:32 UT 
in the 90--160\,\AA\ wavelength range with that of the  RS CVn star V711 Tau, obtained with the EUVE satellite. 
Several of the stronger, highly ionised Fe lines present in these spectra are marked.}

\end{figure*}

In the present paper we significantly expand upon the work of \citet{milligan12} to establish if other high temperature ($>$10\,MK) 
Fe lines apart from those of Fe\,{\sc xxi} can be used to derived plasma densities under flare conditions. Specifically, 
we undertake a comparison of EVE and EUVE observations, plus published results for well-diagnosed high density tokamak plasmas, 
to test the validity of additional highly ionized Fe line ratios present in EVE spectra between 90--160\,\AA\ as 
reasonable density diagnostics for flaring plasmas. More importantly, 
we also assess if they can be employed to derive densities in the high N$_{e}$ ($\simeq$\,10$^{12}$--10$^{13}$\,cm$^{-3}$) regime.

\section{Observational data} 

The solar observations considered in the present paper consist of EVE spectra for two X-class solar flares obtained at event peak, namely that of 2011 February 15 at 01:55:32 UT (X2.2 class) and 2011 August 9 at 08:05:07 UT (X6.9 class). These flares were selected to 
maximise both the intensities of the highly ionized Fe  lines of interest (and hence minimise the effect of blending from low temperature transitions), and the electron density of the emitting plasma (ideally reaching densities close to 10$^{13}$\,cm$^{-3}$). 
Details of the EVE instrument and data reduction procedures may be found in \citet{woods12} and \citet{milligan12}. Briefly, the 
Multiple EUV Grating Spectrograph (MEGS)-A of EVE obtains full-disk spectra of the Sun in the 65--370\,\AA\
wavelength range, at a resolution of $\sim$\,1.0\,\AA, every 10\,s. We have employed the most recent release of the EVE data (Level 2, Version 5), based on new rocket calibration flights, which improves the accuracy 
in the irradiance values presented in earlier releases (\citeauthor{milligan12}). 

Our paper is focused on the $\sim$\,90--160\,\AA\ region of MEGS-A,
which contained numerous emission lines arising from transitions in Fe\,{\sc xviii} -- Fe\,{\sc xxiii}. 
\citet{delzannawoods13} have previously undertaken an analysis of Fe\,{\sc xviii} -- Fe\,{\sc xxiv}
features in EVE spectra, to investigate which of these may be relatively free of blending in solar flares
and hence have the potential to be employed as diagnostics. However, these authors did not consider 
electron density diagnostics in detail, in particular the consistency of densities derived from 
different line ratios. Furthermore, they focused on plasmas with N$_{e}$ in the range $\sim$\,10$^{11}$--10$^{12}$\,cm$^{-3}$, where 
many of the Fe line diagnostics are not particularly useful. For example, the Fe\,{\sc xxi} (142.14 + 142.28)/128.75 ratio recommended
by \citet{milligan12} only varies by $\sim$\,35\%\ between 10$^{11}$ and 10$^{12}$\,cm$^{-3}$. The Fe\,{\sc xxi} 145.73/128.75 ratio recommended
by both \citeauthor{milligan12} and \citeauthor{delzannawoods13} varies by a much greater amount over this density interval (a factor of 
4.8), but with the problem that the 145.73\,\AA\ line is very weak at such values of N$_{e}$, being predicted to be typically 
only 5\%\ of the intensity of 128.75\,\AA. By contrast, our work is primarily concerned with 
the identification and assessment of
diagnostics for much higher density flare plasmas, ideally up to N$_{e}$ $\simeq$ 10$^{13}$\,cm$^{-3}$ or even greater.

In Table 1 we summarise the lines judged by \citet{delzannawoods13}
to be blend free and which may provide electron density sensitive intensity ratios. However, 
in addition we  
include the Fe\,{\sc xxi} 123.83 and Fe\,{\sc xxii} 156.02\,\AA\ transitions. The former was not considered by \citeauthor{delzannawoods13}, while in the case of Fe\,{\sc xxii} 156.02\,\AA\ these authors 
noted that the line appeared to be blended as it implied a very high value of electron density.
To investigate the potential blending of these features, we have calculated  
synthetic spectra with the latest version (8.0.1) of the CHIANTI collisional-radiative modelling packages, appropriate for
plasmas with densities up to $\sim$\,10$^{15}$\,cm$^{-3}$ \citep{dere97,delzanna15}. Spectra have been
generated for electron densities in the range 10$^{11}$--10$^{13}$\,cm$^{-3}$,
and for the standard CHIANTI flare differential emission measure (DEM) distribution. Although each solar flare will have its own characteristics (including DEM), we note that the highly ionized Fe lines are formed over a narrow range of temperatures of maximum fractional abundance, log T$_{max}$ = 7.0--7.1 \citep{bryans09}. The DEMs of different flares do not vary significantly over this temperature interval, as shown by for example Figure 5 of \citet{kennedy13}. Our CHIANTI synthetic spectra 
indicate that blending species (primarily Cr\,{\sc xvii} 122.97, Mn\,{\sc xxi} 124.08 and Cr\,{\sc xx} 156.02\,\AA) should make at 
most $\sim$\,25\%\ 
contributions to the total measured line fluxes of both the Fe\,{\sc xxi} 123.83 and Fe\,{\sc xxii} 156.02\,\AA\ lines.

\begin{table}
\centering
\caption{Highly ionized Fe lines considered in the present analysis}
\label{}
\begin{tabular}{lcc}
\hline
Species & Transition & Wavelength (\AA)
\\
\hline
Fe\,{\sc xx} & 2s$^{2}$2p$^{3}$ $^{2}$D$_{5/2}$ --  2s2p$^{4}$ $^{2}$D$_{5/2}$ & 113.35
\\
Fe\,{\sc xx} & 2s$^{2}$2p$^{3}$ $^{4}$S$_{3/2}$ --  2s2p$^{4}$ $^{4}$P$_{3/2}$ & 121.85
\\
Fe\,{\sc xxi} & 2s$^{2}$2p$^{2}$ $^{3}$P$_{2}$ --  2s2p$^{3}$ $^{3}$P$_{1}$ & 123.83
\\
Fe\,{\sc xxi} & 2s$^{2}$2p$^{2}$ $^{3}$P$_{0}$ --  2s2p$^{3}$ $^{3}$D$_{1}$ & 128.75
\\
Fe\,{\sc xxi} & 2s$^{2}$2p$^{2}$ $^{3}$P$_{1}$ --  2s2p$^{3}$ $^{3}$D$_{1,2}$ & 142.14 + 142.28
\\
Fe\,{\sc xxi} & 2s$^{2}$2p$^{2}$ $^{3}$P$_{2}$ --  2s2p$^{3}$ $^{3}$D$_{3}$ & 145.73
\\
Fe\,{\sc xxii} & 2s$^{2}$2p $^{2}$P$_{3/2}$ --  2s2p$^{2}$ $^{2}$P$_{3/2}$ & 114.41
\\
Fe\,{\sc xxii} & 2s$^{2}$2p $^{2}$P$_{1/2}$ --  2s2p$^{2}$ $^{2}$P$_{1/2}$ & 117.15
\\
Fe\,{\sc xxii} & 2s$^{2}$2p $^{2}$P$_{1/2}$ --  2s2p$^{2}$ $^{2}$D$_{3/2}$ & 135.79
\\
Fe\,{\sc xxii} & 2s$^{2}$2p $^{2}$P$_{3/2}$ --  2s2p$^{2}$ $^{2}$D$_{5/2}$ & 156.02
\\
\hline
\end{tabular}
\end{table}

For late-type stellar observations, we have employed spectra from the EUVE satellite mission, which operated from 1992 June 7 to 2001 
January 31. A full description of the EUVE spectrometers may be found in \citet{abbott96} and references therein. Briefly,
EUVE observed sources over the 70--760\,\AA\ wavelength range using three slitless, imaging spectrometers, with that of relevance to the 
present paper being the short-wavelength (SW) instrument which covered 70--190\,\AA\ at a resolution of $\sim$\,0.5\,\AA. An atlas of 
the spectra for 95 stellar sources from the EUVE Public Archive was published in \citet{craig97}, which also contains details
of the data reduction procedures. We have selected a sample of late-type stars for our study which are extremely 
active and hence show strong emission from highly ionized Fe lines, as well as large electron density estimates
(N$_{e}$ $>$ 10$^{12}$\,cm$^{-3}$) for the emitting coronal plasma. Data for these objects were obtained
from the Mikulski Archive for Space Telescopes (MAST).
The archival spectra were processed with EUVE IRAF software version 1.9 with reference data EGODATA 1.17, and analyzed using IRAF and custom IDL routines. To improve the signal-to-noise of the EUVE data, 
we created new summed datasets using all available SW observations in MAST for UX Ari (4 spectra), 
V711 Tau (7) and AD Leo (9), while for 44i Boo and AU Mic we analysed existing co-added data. The total exposure times of 
the summed EUVE spectra are listed in Table 2. Detailed information on each star and the relevant EUVE spectral datasets may be found in the
following references: \citet{brickhousedupree98} for 44i Boo, \citet{monsignorifossi96} for AU Mic,
\citet{sanzforcada02} for UX Ari and V711 Tau, and \citet{micela02} for AD Leo.

\begin{landscape}

\begin{table}

\caption{Fe ion line intensity ratios$^{a}$ and derived logarithmic electron densities}
\label{}
\begin{tabular}{lccccccccccc}
\hline
Line ratio & 2011 Feb 15 flare & 2011 Aug 9 flare & 44i Boo & AU Mic & UX Ari
& V711 Tau & AD Leo & FTU & PLT & PLT & PLT
\\
& 01:55:32 UT & 08:05:07 UT & 141 ks$^{b}$ & 69.6 ks$^{b}$ & 450 ks$^{b}$ 
& 790 ks$^{b}$ & 1100 ks$^{b}$ & tokamak$^{c}$ & tokamak$^{d}$ & tokamak$^{d}$ & tokamak$^{e}$
\\
\hline
Fe\,{sc xx} 113.35/121.85 & \ldots & 0.20$\pm$0.02 & 0.21$\pm$0.12 & 0.23$\pm$0.14 & 0.16$\pm$0.10 & 0.18$\pm$0.12 & 0.25$\pm$0.17 & \ldots & \ldots & \ldots & \ldots
\\
& \ldots & 12.6$\pm$0.1 & 12.7$^{+0.5}_{-0.9}$ & 12.8$^{+0.5}_{-1.0}$ & 12.4$^{+0.5}_{-1.2}$ & 12.5$^{+0.6}_{-1.3}$ 
& 12.9$^{+0.6}_{-1.2}$ &  \ldots  &\ldots & \ldots & \ldots
\\
Fe\,{\sc xxi} (142.14+142.28)/128.75 & 0.13$\pm$0.02 & 0.16$\pm$0.02 & 0.51$\pm$0.24 & 0.21$\pm$0.11 & 0.29$\pm$0.15 & 0.17$\pm$0.07 
& 0.38$\pm$0.17 & 
1.3$\pm$0.2 & \ldots & \ldots & 0.91
\\
& 11.7$^{+0.2}_{-0.6}$ & 12.0$^{+0.1}_{-0.2}$ & 12.9$^{+0.3}_{-0.4}$ &  12.3$^{+0.3}_{-1.7}$ & 12.5$^{+0.3}_{-0.7}$ 
& 12.0$^{+0.4}_{-1.4}$ & 12.7$^{+0.3}_{-0.4}$ & 13.8$^{+0.3}_{-0.2}$ & \ldots & \ldots & 13.4
\\
Fe\,{\sc xxi} 145.73/128.75 & 0.21$\pm$0.04 & 0.17$\pm$0.03 & 0.34$\pm$0.19 & 0.22$\pm$0.06 & 0.11$\pm$0.09 & 0.05$\pm$0.02 
& 0.06$\pm$0.03 & 1.0$\pm$0.2 & \ldots & \ldots & 0.61
\\
& 12.4$^{+0.1}_{-0.2}$ & 12.2$\pm$0.1 & 12.7$^{+0.4}_{-0.6}$  & 12.4$^{+0.2}_{-0.2}$ & 12.0$^{+0.3}_{-1.2}$
& 11.5$^{+0.2}_{-0.4}$ & 11.6$^{+0.2}_{-0.5}$ & 13.7$^{+0.3}_{-0.3}$ & \ldots & \ldots
& 13.2
\\
Fe\,{\sc xxi} 123.83/(142.14+142.28) & 0.91$\pm$0.17 & 0.49$\pm$0.06 & 0.37$\pm$0.08 & 0.64$\pm$0.41 & 0.50$\pm$0.37 
& 0.20$\pm$0.11 & 0.28$\pm$0.19 & \ldots & \ldots & \ldots & \ldots
\\
& L$^{f}$ & L$^{f}$ & $<$\,10.0 & $<$\,11.8 & $<$\,12.7 & 12.1$^{-\infty}_{+1.6}$ 
& 10.9$^{-\infty}_{+2.8}$ & \ldots & \ldots & \ldots &\ldots
\\
Fe\,{\sc xxii} 114.41/135.79 & 0.25$\pm$0.03 & 0.25$\pm$0.02 & 0.41$\pm$0.23 & 0.26$\pm$0.09 & 0.29$\pm$0.11 
& 0.21$\pm$0.04 & 0.30$\pm$0.08 & \ldots & 0.79 & 0.60 & 0.82
\\
& 12.5$^{+0.2}_{-0.3}$ & 12.5$^{+0.2}_{-0.1}$ & 13.1$^{+0.4}_{-\infty}$ & 12.6$^{+0.4}_{-\infty}$ & 12.8$^{+0.3}_{-\infty}$ 
& 12.1$^{+0.4}_{-\infty}$  & 12.8$^{+0.3}_{-0.6}$ & \ldots & 13.7 & 13.5 & 13.8
\\
Fe\,{\sc xxii} 114.41/117.15 & 0.17$\pm$0.02 & 0.25$\pm$0.02 & 0.54$\pm$0.30  & 0.22$\pm$0.08 & 0.23$\pm$0.07 & 0.23$\pm$0.07 
& 0.24$\pm$0.04 & 1.1$\pm$0.3 & 0.67 & 0.33 & 0.40
\\
& $<$\,11.4 & 12.6$^{+0.1}_{-0.2}$ & 13.4$^{+0.4}_{-0.9}$ & 12.3$^{+0.5}_{-\infty}$ & 12.4$^{+0.4}_{-\infty}$ 
& 12.4$^{+0.4}_{-\infty}$ & 12.5$^{+0.2}_{-0.6}$ & 14.1$^{+0.5}_{-0.3}$ & 13.6 & 12.9 & 13.1
\\
Fe\,{\sc xxii} 156.02/135.79 & 0.14$\pm$0.04 & 0.076$\pm$0.013 & 0.18$\pm$0.14  & 0.29$\pm$0.07 & 0.19$\pm$0.12
& 0.19$\pm$0.16 & 0.21$\pm$0.10 & 0.55$\pm$0.06 & \ldots & \ldots & 0.59
\\
& 12.9$^{+0.1}_{-0.3}$ & 12.4$^{+0.1}_{-0.2}$ & 13.0$^{+0.4}_{-\infty}$ & 13.4$^{+0.1}_{-0.2}$ & 13.1$^{+0.3}_{-0.8}$
& 13.1$^{+0.2}_{-0.3}$ & 13.1$^{+0.3}_{-0.4}$ & 13.8$^{+0.1}_{-0.1}$ & \ldots 
& \ldots & 13.9
\\
Mean density & 12.4$\pm$0.5  & 12.4$\pm$0.2 & 13.0$\pm$0.3 & 12.6$\pm$0.4 & 12.5$\pm$0.4 & 12.3$\pm$0.5 & 12.6$\pm$0.5 & 
13.9$\pm$0.2 & 13.7$\pm$0.1 & 13.2$\pm$0.4 & 13.5$\pm$0.4
\\
Log N$_{e}$(other)$^{g}$ & 
\ldots  &  
\ldots& 13.2$\pm$0.7
& 12.7$\pm$0.5  
& 12.5$\pm$0.4 & 12.2$\pm$0.1 & 12.9$\pm$0.3   & 13.9 & 13.5 & 12.7 & 13.4
\\
\hline
\end{tabular}

$^{a}$Ratios are in photon units.
\\
$^{b}$Total exposure time of the summed EUVE short-wavelength spectrum.
\\
$^{c}$From \citet{fournier01}.
\\
$^{d}$From \citet{stratton84}.
\\
$^{e}$From \citet{stratton85}.
\\
$^{f}$L indicates that the observed ratio is greater than the theoretical low density limit.
\\
$^{g}$Stellar electron densities derived from a range of high temperature ions in the EUVE spectra.
The tokamak values are those measured by \citet{fournier01} or \citet{stratton84,stratton85}.

\end{table}

\end{landscape}

The final sets of EUV spectra considered here are from the Frascati Tokamak Upgrade (FTU) and Princeton Large Torus (PLT) tokamaks, obtained by \citet{fournier01} and \citet{stratton84,stratton85}, respectively. All three datasets have a spectral resolution of
$\sim$\,0.7\,\AA, similar to those of the EVE and EUVE observations. The tokamak spectra will not show the same level
of blending as the astrophysical data, due to the latter sampling a larger number of elements. However, they should still 
provide a useful comparison with the solar/stellar observations, as most of the significant 
emission lines in the EUV region of interest arise from other high ionization Fe transitions. 
This is shown by, for example, a comparison of the Fe spectrum of the FTU tokamak between 85--140\,\AA\ in Figure 2 of 
\citeauthor{fournier01} with the EVE and EUVE data in Figure 1, which appear very similar.
Hence the tokamak spectra should allow the reliability and consistency of highly ionized Fe line diagnostics to be assessed. Tokamaks
are also particularly useful to the present study as their plasmas have very large values of N$_{e}$, similar to or greater than those of high electron density flares.

\section{Results and discussion} 

Line fluxes for the Fe ion lines in both the EVE and EUVE spectra were determined by fitting multiple Gaussian profiles
to the observations, with some example fits shown in Figures 2 and 3. The resultant line intensity ratios (in photon units), along with the $\pm$\,1$\sigma$ errors, are listed in Table 2. Also included in the table are the observed line ratios for the FTU and PLT
tokamak spectra, taken directly from \citet{fournier01} and \citet{stratton84,stratton85}.

\begin{figure*}
		\includegraphics[width=14cm]{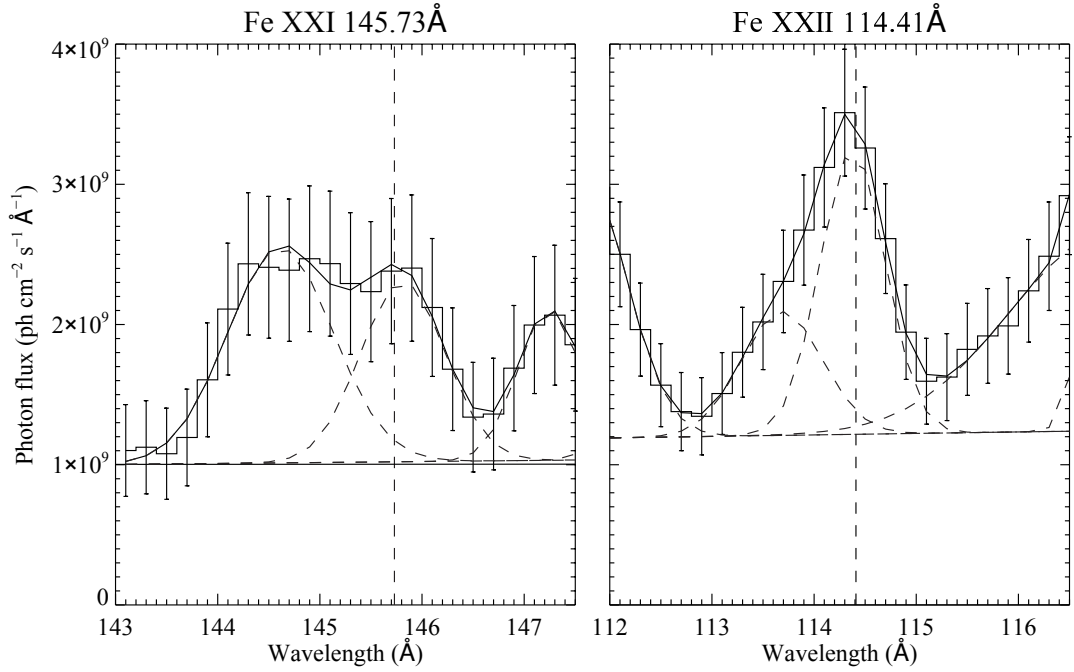}
    \caption{Portions of the EVE spectrum for the 2015 February 15 solar flare at 01:55:32 UT, showing the Fe\,{\sc xxi}
    145.73 and 
Fe\,{\sc xxii} 114.41\,\AA\ emission features, both marked with vertical dashed lines. Also shown with dashed lines 
are the multi-line fits to the profiles, which include emission features due to Ca\,{\sc xv} 144.31 + Fe\,{\sc xxiii} 144.39\,\AA, 
Fe\,{\sc xxiii} 147.25\,\AA,
Fe\,{\sc xxi} 113.29 + Fe\,{\sc xx} 113.35\,\AA, plus the wings of Fe\,{\sc xix} 111.70 + Ni\,{\sc xxiii} 111.83\,\AA, and 
Fe\,{\sc xxii} 117.15\,\AA.}
\end{figure*}

\begin{figure*}
		\includegraphics[width=10cm]{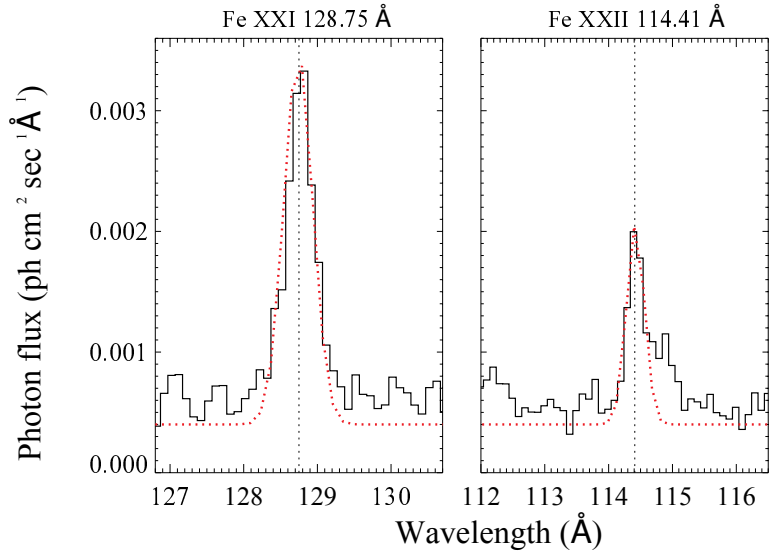}
    \caption{Portions of the EUVE Short Wavelength (SW) spectrometer observations 
for the RS CVn star V711 Tau, with a total exposure time of 790\,ks, 
obtained during the period 1992 October 22 to 1999 September 13. Profile fits to the Fe\,{\sc xxi} 128.75\,\AA\ and 
Fe\,{\sc xxii} 114.41\,\AA\ emission features are shown as dashed lines.}
\end{figure*}

We have derived values of electron density from the observed  EVE, EUVE and tokamak line ratios using theoretical results from the latest version (8.0.1) of CHIANTI
\citep{dere97,delzanna15}. All of the line ratio calculations were performed at the value of 
T$_{max}$  of the relevant Fe ion \citep{bryans09}, although we note that the results are not sensitive to the adopted temperature.
For example, changing the electron temperature for the Fe\,{\sc xxi}  (142.14 + 142.28)/128.75 
line ratio calculations from T$_{max}$ = 1.1$\times$10$^{7}$\,K to 2$\times$10$^{7}$\,K leads to  a $<$\,0.1 dex variation in the derived 
value of N$_{e}$. Similarly, for Fe\,{\sc xxii} 114.41/135.79, adopting T$_{e}$ = 2.5$\times$10$^{7}$\,K rather than T$_{max}$
= 1.3$\times$10$^{7}$\,K also results in a $<$\,0.1 dex change in the density estimate. 
Hence even if the plasma is not in ionization equilibrium and the temperature is significantly different from T$_{max}$, as
may be the case during a flare (see, for example, \citealt{kawate16}), this should not affect the derived values of N$_{e}$. Furthermore, the timescale to 
achieve ionization equilibrium scales with electron density, and for the high N$_{e}$ events considered here the ionization
state of Fe should be close to equilibrium \citep{bradshaw09}. 

Also listed in Table 2 are the electron densities, denoted N$_{e}$(other), determined from a range of 
high temperature ions in the EUVE spectra, in particular those with 
lines known to be blended in the EVE spectra \citep{delzannawoods13} but which should be relatively unblended in the (somewhat) greater resolution EUVE observations. These include density diagnostics such as Fe\,{\sc xx} 121.85/110.63, Fe\,{\sc xxi} 121.21/128.75 and 
Fe\,{\sc xxii} 116.27/117.15. 
For
the tokamak observations, N$_{e}$(other) is the experimental plasma density, determined from interferometry measurements to an accuracy of approximately
$\pm$\,15\%\ \citep{stratton84,stratton85,fournier01}. 

The first point to note from an inspection of Table 2 is that the Fe\,{\sc xxi} 123.83\,\AA\ line is clearly blended to
a very significant degree in the EVE observations, as the measured values of Fe\,{\sc xxi} 123.83/(142.14 + 142.28) are both far greater
(by factors of 1.7 and 3.1) than the theoretical low density limit of 0.29. As noted in Section 2, this line was not considered
by \citet{delzannawoods13}  in their assessment of EVE flare data, although 
it is not predicted to be severely blended. However, blending 
species must make a much larger contribution to the total intensity than evisaged, including perhaps unidentified transitions.  

For the other Fe ion lines in the EVE spectra, the resultant intensity ratios imply electron densities that are reasonably consistent,
with discrepancies that average only 0.3 dex from the mean value of log N$_{e}$ = 12.4 for both flares. 
Similarly, for the EUVE datasets, the line ratios lead to electron density estimates
for each star that are consistent within the errors, 
and indicate average densities which show no significant discrepancies with those derived from
a wider range of high temperature N$_{e}$-diagnostic lines. Differences between the mean densities and the values of N$_{e}$(other)
average only 0.1 dex, and do not exceed 0.3 dex. We note that as the EUVE spectra analysed here are the sum of several SW
datasets, the derived densities will not represent specific flare values, but rather time-averaged estimates over the duration of the 
observations. However, this is not an issue as we are only concerned with the consistency of densities derived from different line ratios, and hence a comparison of time-averaged values is appropriate. In addition, it is worth pointing out that the ratio values for individual flares are actually very similar to those from the summed spectra. For example, for AU Mic 
the measured Fe\,{\sc xxii} 114.41/135.79 and 114.41/117.15 ratios in Table 2 are 0.26 and 0.22, respectively, while those for
a specific flare (interval b of \citealt{monsignorifossi96}) are 0.30 and 0.23. The similar ratio values reflect the fact that the spectral emission
is dominated, as might be expected, by high density flaring plasma. However, we employ summed EUVE spectra in our work as these provide
better signal-to-noise and hence more reliable detections, especially for weak features.

In the case of the tokamak spectra, the derived average values of N$_{e}$ are consistent with the
experimental data in three out of four instances, with discrepancies of less than 0.2 dex. The exception is the PLT data of 
\citet{stratton84} at an experimental density of log N$_{e}$ = 12.7, where the diagnostics indicate
log N$_{e}$ = 13.2$\pm$0.4. This is due to the Fe\,{\sc xxii} 114.41/135.79 ratio being much larger 
than expected; the measured value is 0.60, while for an experimental density of log N$_{e}$ = 12.7 the theoretical ratio is only
0.28. \citeauthor{stratton84} also noted this discrepancy, and it
remains unexplained. However, the problem must lie with the 135.79\,\AA\ line measurement 
in the spectrum, as the Fe\,{\sc xxii} 114.41/117.15 ratio shows no significant differences between theory (0.27 at log N$_{e}$ =
12.7) and 
observation (0.33), and hence the 114.41\,\AA\ experimental intensity should be secure. 
We note that there is no issue with the 135.79\,\AA\ line in the FTU
spectrum of \citet{fournier01}, with good agreement between the density derived from 
Fe\,{\sc xxii} 156.02/135.79 (log N$_{e}$ = 13.8$\pm$0.1) and the measured value (log N$_{e}$ = 13.9).
Furthermore, there is consistency between theory and observation for 
Fe\,{\sc xxii} 114.41/135.79 in the 
PLT spectrum for the log N$_{e}$ = 13.5 plasma, with the measured value indicating log N$_{e}$ = 13.7,
providing support for the reliability of the CHIANTI line ratio calculations.

\section{Conclusions}

The consistency of the electron densities determined from the EVE observations, combined with that 
also found from the EUVE and tokamak data of similar spectral resolution, indicates that the 
line ratios in Table 2 with the exception of Fe\,{\sc xxi} 123.83/(142.14 + 142.28) should provide reasonable
estimates of N$_{e}$ for the high temperature ($\sim$\,10\,MK)  plasma in solar flares. More importantly,
the EUVE and tokamak
results provide 
support for the line ratio diagnostics being reliable up to very high values of N$_{e}$ ($\simeq$\,10$^{13}$\,cm$^{-3}$), 
and confirm the large densities derived for the solar events. \citet{milligan12} found that two of the 
Fe\,{\sc xxi} ratios in Table 2 --- 145.73/128.75 and 
(142.14 + 142.28)/128.75 --- provided good estimates of density at N$_{e}$ $\simeq$ 10$^{12}$\,cm$^{-3}$. (We note in passing that these authors also considered Fe\,{\sc xxi} 121.21/128.75, which we exclude as \citet{delzannawoods13} point out 
that the 121.21\,\AA\ feature
is very weak and in the wing of the strong Fe\,{\sc xx} 121.84\,\AA\ line). However, the present work significantly
extends the
\citeauthor{milligan12} analysis 
to show that four additional ratios of Fe\,{\sc xx} and Fe\,{\sc xxii} may be confidently used as diagnostics, even at the low
spectral resolution of EVE, and furthermore allow 
a high electron density 
regime to be reliably investigated. As EVE obtains full-disk spectra at a very high cadence of 10\,s (as opposed to
its low spectral resolution), this in turn indicates that the observations may be employed for reliable
temporally-resolved studies of flare densities in the high N$_{e}$ regime ($\sim$\,10$^{12}$--10$^{13}$\,cm$^{-3}$), 
and in particular the investigation of
the maximum density that can be achieved during flare peak. The EVE instrument has observed hundreds of flares during its 4 years of operation, providing a very large dataset for such studies. Measurement of as many as possible 
of the line ratios in Table 2 would be preferred, as the average densities are in better agreement with N$_{e}$(other) than the 
individual values, which show a large amount of scatter. However, the best diagnostics --- based on their density sensitivity particularly at high N$_{e}$, and how well the derived densities agree with N$_{e}$(other) ---
are probably Fe\,{\sc xx} 113.35/121.85,  Fe\,{\sc xxi} (142.14 + 142.28)/128.75 and 
Fe\,{\sc xxii} 114.41/135.79. 

Now that we have confirmed the usefulness of MEGS-A  data for determining  high flare densities, the next step is to extend our 
analyses to the MEGS-B spectral region (350--1050\,\AA), which samples lower temperature emission lines. The overall aim of this work will be to identify diagnostics which allow time profiles of flare densities simultaneously across a broad temperature range 
($\sim$\,0.1--10\,MK).

\section*{Acknowledgements}

FPK and MM are grateful to the Science and Technology Facilities Council for financial support.
The research leading to these results has received funding from the European Community's Seventh Framework Programme (FP7/2007-2013) under grant agreement no. 606862 (F-CHROMA). We are also grateful to the Leverhulme Trust for financial support 
via grant F/00203/X. ROM acknowledges support from NASA LWS/SDO Data Analysis grant NNX14AE07G.
CHIANTI is a collaborative project involving George Mason University, the University of Michigan (USA) and the University of Cambridge (UK). The EUVE data presented in this paper were obtained from the Mikulski Archive for Space Telescopes (MAST). STScI is operated by the Association of Universities for Research in Astronomy, Inc., under NASA contract NAS5-26555. Support for MAST for non-HST data is provided by the NASA Office of Space Science via grant NNX09AF08G and by other grants and contracts.









\bsp	
\label{lastpage}
\end{document}